\documentclass[twocolumn,twocolappendix]{aastex631}
\usepackage{graphicx}	% Including figure files
\usepackage{amsmath}	% Advanced maths commands
\usepackage{amssymb}	% Extra maths symbols

\shorttitle{P-AIRCARS Implementation for the MWA}
\shortauthors{Kansabanik et al.}

\graphicspath{{./}}

\begin{document}

\title{Tackling the Unique Challenges of Low-frequency Solar Polarimetry with the Square Kilometre Array Low Precursor: Pipeline Implementation}

\author[0000-0001-8801-9635]{Devojyoti Kansabanik}
\affiliation{National Centre for Radio Astrophysics, Tata Institute of Fundamental Research, Pune University Campus, Pune 411007, India}

\author[0000-0002-2864-4110]{Apurba Bera}
\affiliation{International Centre for Radio Astronomy Research, Curtin University, Bentley, WA 6102, Australia}
\affiliation{Inter-University Centre for Astronomy and Astrophysics, Post Bag-4, Ganeshkhind, Pune 411007, India}
\affiliation{National Centre for Radio Astrophysics, Tata Institute of Fundamental Research, Pune University Campus, Pune 411007, India}

\author[0000-0002-4768-9058]{Divya Oberoi}
\affiliation{National Centre for Radio Astrophysics, Tata Institute of Fundamental Research, Pune University Campus, Pune 411007, India}

\author[0000-0002-2325-5298]{Surajit Mondal}
\affiliation{Center for Solar-Terrestrial Research, New Jersey Institute of Technology, 323 M L King Jr Boulevard, Newark, NJ 07102-1982, USA}

\correspondingauthor{Devojyoti Kansabanik}
\email{dkansabanik@ncra.tifr.res.in, devojyoti96@gmail.com}

\accepted{December 16, 2022}

\begin{abstract}
The dynamics and the structure of the solar corona are determined by its magnetic field. Measuring coronal magnetic fields is, however, extremely hard. Polarization of the low-frequency radio emissions has long been recognized as one of the few effective observational probes of magnetic fields in the mid and high corona. However the extreme intrinsic variability of this emission; the limited ability of most existing instrumentation available till recently to capture it; and the technical challenges involved, have all contributed to severely limiting its use.
The high dynamic range spectro-polarimetric snapshot imaging capability necessarily needed for radio coronal magnetography is now within reach.
This has been enabled by the confluence of data from the Murchison Widefield Array (MWA), a Square Kilometre Array (SKA) precursor and our unsupervised and robust polarization calibration and imaging software pipeline dedicated for the Sun -- Polarimetry using Automated Imaging Routine for the Compact Arrays of the Radio Sun (P-AIRCARS).
Here we present the architecture and  implementation details of P-AIRCARS. 
Although, the present implementation of P-AIRCARS is tuned for the MWA, the algorithm itself can easily be adapted for future arrays like the SKA1-Low.
We hope and expect that P-AIRCARS will enable exciting new science with instruments like the MWA, and will encourage wider use of radio imaging in the larger solar physics community. 
\end{abstract}


\section{Introduction}\label{sec: introduction to the problem}
Solar phenomena span an enormous range of time scales, from solar cycle to flares and in terms of energy from the most massive coronal mass ejections (CMEs) to the barely discernible nanoflares. It is now well understood that the solar magnetic field is the primary driver of all of these phenomena. These magnetic fields also couple the solar atmosphere to the solar interior. Hence, to understand coronal physics and dynamics it is essential to measure and understand the ever-evolving coronal magnetic fields. Very recently, \cite{Yang2020} have successfully demonstrated a method of measuring the global coronal magnetic fields in the range $1.05$--$1.35\ \mathrm{R_\odot}$ using near-infrared observations, where $\mathrm{R_\odot}$ is the solar radius. Measuring coronal magnetic fields is also a key objective for the Daniel K. Inouye Solar Telescope \citep[DKIST;][]{Rast2021}, though these measurements are likely to remain limited to less than $1.5\ \mathrm{R_\odot}$.

The observed polarization properties of low-frequency coronal radio emissions can serve as excellent probes of coronal magnetic fields, even at middle and higher coronal heights. This is because the magnetic field affects the polarization of radio emissions arising from the coronal plasma \citep[e.g.][]{Alissandrakis2021}. Polarization observations also enable a detailed understanding of the emission mechanism of these low-frequency coronal radio emissions. Many successful examples exist in the literature, though their numbers have been rather small and these studies have remained limited to comparatively brighter and highly polarized emissions. Most of these studies have relied on dynamic spectra, \citep[e.g.,][etc.] {McLeanBook,hari2014,Kumari2017,Pulupa2020,Ramesh_2022}. In only a handful of instances, either the information of spatial location \citep[e.g.][]{Mercier1990,Morosan2022} and/or spatial structure \citep[e.g.][]{Patrick2019, Rahman2020} of the sources are also available. 

High dynamic range spectro-polarimetric solar imaging studies at low radio frequencies are very rare. Brightness temperature ($T_\mathrm{B}$) of solar radio emission varies by as much as about nine orders of magnitude, and their fractional polarization can vary by about two orders of magnitude \citep{McLeanBook,Kansabanik_principle_AIRCARS}. These emissions can change drastically over short temporal and spectral spans. Very often, faint emissions can simultaneously be present with very bright emissions. This imposes the need for high dynamic-range polarimetric imaging over short temporal and spectral spans. These challenging requirements along with the technical and instrumental limitations at low radio frequencies have severely limited polarimetric solar radio imaging studies, despite their well-appreciated importance.

Radio interferometry is a Fourier synthesis imaging technique. Hence, the fidelity of the detected emission in spectroscopic snapshot radio images depends on the density of the antenna distribution of the array. This essential requirement for high-fidelity spectroscopic snapshot solar imaging is met by one of the new technology instruments, the Murchison Widefield Array \citep[MWA;][]{lonsdale2009,Tingay2013,Wayth2018}. Operating in the 80 to 300 MHz band, the MWA is also a Square Kilometre Array (SKA) \citep{ska_concept} precursor. Though the MWA data are intrinsically capable of yielding high-fidelity solar images, doing so involves surmounting several challenges. These challenges have successfully been dealt with in the robust calibration and total intensity imaging pipeline developed by \citet{Mondal2019}, named \textit{Automated Imaging Routine for Compact Arrays of the Radio Sun} (AIRCARS). The high dynamic range spectroscopic snapshot images it delivers represent the state-of-the-art and have led to several interesting results at low radio frequencies, such as the discovery of quasi-periodic pulsations of solar radio bursts \citep{Mohan2019a,Mohan2019b,Mondal2021a}, spectroscopic imaging study of type-II radio burst \citep[][accepted at A\&A]{Bhunia2022}, and the discovery and investigations of Weak Impulsive Narrowband Quiet Sun Emissions \citep[WINQSEs;][]{Mondal2020b,Mondal2021b,Bawaji2021,mondal2022study} 
and the measurement of plasma parameters of CMEs using gyrosynchrotron spectrum modeling \citep{Mondal2020a}. 

AIRCARS was designed for total intensity imaging. But unlocking the potential of low radio frequency solar science requires polarimetric imaging.
%, as argued at the beginning of this section. 
An algorithm, \textit{Polarimetry using Automated Imaging Routine for the Compact Arrays of the Radio Sun} (P-AIRCARS) \footnote{Documentation available online at \url{https://p-aircars.readthedocs.io/en/latest}}, has been developed to achieve this and was presented in \citet{Kansabanik2022_paircarsI} (Paper-I hereafter). In addition to providing the functionality for polarimetric imaging, P-AIRCARS also offers several improvements over AIRCARS including a modular architecture, an improved and flexible calibration strategy, and more efficient parallelization. This companion paper describes the implementation and architecture of P-AIRCARS. 

We organize the paper as follows. We first discuss the design principles of P-AIRCARS in Section \ref{sec:key_principles}. We briefly describe the calibration algorithm in Section \ref{sec:paircars_cal_algorithm}. Section \ref{sec:pipeline_arch} describes the architecture of the pipeline. A custom-developed flagging module for P-AIRCARS is described in Section \ref{sec:ankflag}. Section \ref{sec:determine_params} describes the choices of parameters for calibration and imaging followed by salient features of P-AIRCARS in Section \ref{sec:features}. 
Aspects related to hardware and software requirements for P-AIRCARS and its performance are discussed in Section \ref{sec:requirement}. We discuss current limitations and future works in Section \ref{sec:limitation_and_future} and ends with the conclusions in Section \ref{sec:conclusion}.

\section{Design Principles of P-AIRCARS}\label{sec:key_principles}
The instantaneous bandwidth of the MWA is 30.72 MHz, which can be split into 24 {\em coarse channels} of 1.28 MHz each and distributed across the entire band. 
MWA solar observations are typically done with 10 kHz and 0.5 s resolution. Making images at this temporal and spectral resolution over the useful part of the complete band leads to approximately 370,000 images for an observing duration of 4 minutes. In the next phase of the MWA, the data volume will dramatically increase. The future SKA is expected to produce even larger volumes of data. Performing the calibration and snapshot spectro-polarimetric imaging of such large volumes of data manually is infeasible. One necessarily needs a software pipeline, ideally with the following capabilities :
\begin{enumerate}
    \item The calibration and imaging algorithm and implementation must be robust.
    \item It should be capable of unsupervised operation.
    \item The algorithms it implements should be data-driven and not rely on ad-hoc assumptions.
    \item The software implementation should provide efficient parallelization which scales well with the available hardware resources.
\end{enumerate}

While AIRCARS provided state-of-the-art total intensity images, it did not meet the last two of the requirements stated above. AIRCARS made some ad-hoc assumptions while choosing the calibration and imaging parameters, and the calibration approach limited the parallelization only across the frequency axis. Hence, adding polarization calibration alone to the AIRCARS framework is not sufficient. We take this opportunity of adding polarimetric calibration to completely redesign the software framework and the calibration approach to overcome these limitations. In addition, P-AIRCARS has also been developed to be deployable across a variety of hardware environments -- ranging from laptops and workstations to high-performance computers (HPCs). This makes it very flexible. 

Radio interferometric imaging inherently involves a steep learning curve.
A consequence has been that solar radio imaging has been the domain of a comparatively small number of expert practitioners and has not found widespread adoption in solar physics, as compared to other wavebands. One of the objectives for P-AIRCARS is to overcome this barrier, especially in view of the new generation of much more capable radio instrumentation becoming available.
To achieve this, P-AIRCARS has been designed to work without requiring any radio interferometry-specific input from the user. 

As a corollary of the above requirement, P-AIRCARS is designed to be fault-tolerant, in the sense that when it encounters issues, it makes smart decisions about updating the parameters for calibration and imaging based on the nature of the issue faced. For a well-informed user, P-AIRCARS allows complete flexibility to tune the algorithms as desired. The rest of the paper describes the software framework, calibration, and parallelization strategies adopted for P-AIRACRS following the design principles described here.

\section{Brief Description of the Calibration Algorithm}\label{sec:paircars_cal_algorithm}
We have implemented a robust polarization calibration algorithm \citep{Kansabanik2022_paircarsI} in P-AIRCARS, developed based on the Measurement Equation framework \citep{Hamaker1996_1,Hamaker2000}. Being an aperture array instrument, the MWA has a large field of view (FoV), and high primary beam sidelobes \citep{neben2015,Sokowlski2017,Line2018}. Hence, at the MWA, calibrator observations during the daytime are contaminated by solar emissions. P-AIRCARS implements a self-calibration-based calibration algorithm, which uses some well-known properties of the low-frequency quiet solar emissions \citep{Kansabanik2022_paircarsI} along with the primary beam response of the MWA antenna tiles \citep{Sokowlski2017}. This algorithm is described in detail in paper-I. Here we describe it briefly to place the implementation details in context. 

An interferometer measures the cross-correlations between its antenna pairs. The measured cross-correlation ({\it visibility}), $V_\mathrm{ij}^\prime$, between antennas $\mathrm{i}$ and $\mathrm{j}$ can be expressed in terms of its true value, $V_\mathrm{ij}$ through the {\it Measurement Equation} \citep{Hamaker2000},
\begin{equation}\label{eq:measurement_eq}
    V_\mathrm{ij}^\prime=J_\mathrm{i}\ V_\mathrm{ij}\ J_\mathrm{j}^\dagger + N_\mathrm{ij}
\end{equation}
where, $J_\mathrm{i}$s are the $2\times2$ Jones matrices representing the antenna-dependent instrumental and atmospheric propagation effects, and $N_\mathrm{ij}$ is the additive noise of the instrument. It is standard practice in interferometry to break $J_\mathrm{i}$s into multiple terms, each describing a particular instrumental and/or atmospheric propagation effect. At low radio frequencies, the ionospheric propagation effect is the only major atmospheric effect. We decompose $J_\mathrm{i}$s as:
\begin{equation}\label{eq:jones_terms}
\begin{split}
    J_\mathrm{i}(\nu,\ t,\ \Vec{l})&=G_\mathrm{i}(t)\ B_\mathrm{i}(\nu)\ K_\mathrm{{cross}}(\nu,\ t)\ D_\mathrm{i}(\nu,\ t)\\ 
    &\times E_\mathrm{i}(\nu,\ t,\ \vec{l})
\end{split}
\end{equation}
where, $\nu,\ t$ and $\vec{l}$ represent frequency, time, and direction in the sky plane. These individual terms in Equation \ref{eq:jones_terms} for antenna $\mathrm{i}$ are:
\begin{enumerate}
    \item $G_\mathrm{i}(t)$ : Product of time-dependent instrumental and ionospheric gain.
    \item $B_\mathrm{i}(\nu)$ : Instrumental bandpass.
    \item $K_\mathrm{cross}(\nu,\ t)$: Phase difference between the receptors for the two orthogonal polarization (X and Y, in the case of MWA) for the reference antenna. This is also referred to as the cross-hand phase.
    \item $E_\mathrm{i}(\nu,\ t,\ \vec{l})$ : Direction dependent primary beam model.
    \item $D_\mathrm{i}(\nu,\ t)$ : Direction independent error on the primary beam model.
\end{enumerate}

To obtain $V_\mathrm{ij}$ from $V_\mathrm{ij}^\prime$, each of these terms needs to be estimated precisely and corrected for. They are estimated in three major calibration steps:
\begin{enumerate}
    \item {\bf Intensity self-calibration : } 
    $\mathrm{G_i}(t)$s are estimated and corrected in this step following the approach detailed in \citet{Mondal2019} and \citet{Kansabanik_principle_AIRCARS}.
    \item {\bf Bandpass self-calibration : } This step estimates and corrects for $B_\mathrm{i}(\nu)$s over each of the 1.28 $\mathrm{MHz}$ coarse channels. Data from quiet solar times are used for this and the integrated solar flux density is assumed to remain constant across a coarse channel.
    \item {\bf Polarization calibration : } This involves first correcting for $K_\mathrm{cross}(\nu,\ t)$, $E_\mathrm{i}(\nu,\ t,\ \vec{l})$ which are estimated independently. Next the $D_\mathrm{i}s$ are estimated and corrected using a perturbative self-calibration-based algorithm described in Paper-I.
\end{enumerate}
These three calibration steps form the three main pillars of the full Jones calibration algorithm of P-AIRCARS. 

\section{Architecture of the Pipeline}\label{sec:pipeline_arch}
\begin{figure}[!t]
    \centering
    \includegraphics[trim={1.5cm 2.5cm 5cm 1cm},clip,scale=0.36]{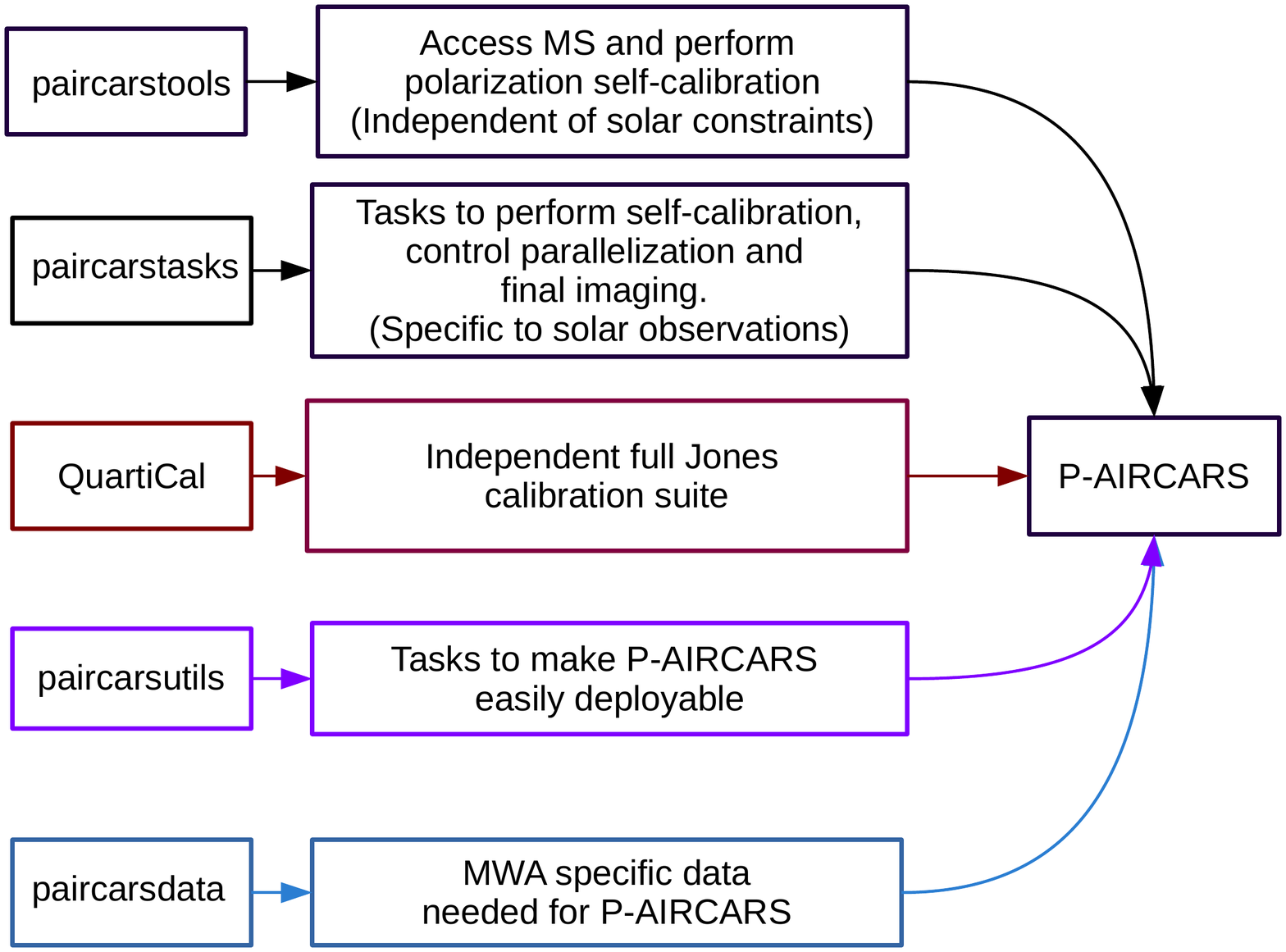}
    \caption{{\textbf{Schematic diagram of P-AIRCARS highlighting its main modules.}}}
    \label{fig:module_schematic}
\end{figure}

P-AIRCARS architecture is highly modular. It has been written with ease of maintenance and adoption to other interferometers with compact core configurations in mind. A large fraction of the P-AIRCARS is written in {\em Python 3}. Some of its core modules used for calibration and flagging are written in {\em C/C++}. A schematic diagram of P-AIRCARS describing all of its modules is shown in Figure \ref{fig:module_schematic}.

The two core modules of P-AIRCARS are \textsf{paircarstools} and \textsf{paircarstasks} and are shown by black boxes in Figure \ref{fig:module_schematic}. \textsf{paircarstools} contains functionalities to perform full polarization self-calibration
without imposing any constraint(s) specific to solar observation and/or the interferometer used. The optimization specific to solar observing is done by \textsf{paircarstasks}, which uses the functionality provided by \textsf{paircarstools}.

The third module, \textsf{QuartiCal}{\footnote{\url{https://pypi.org/project/quartical/}}} is a successor of the full Jones calibration software suite, \textsf{CubiCal} \citep{cubical2018,Cubical_robust2019}.
The \textsf{paircarsutils} module provides the utilities for the deployment of P-AIRCARS across a range of hardware and software architectures and its efficient parallelization. 
The \textsf{paircarsdata} module provides a collection of information specific to the MWA (e.g. the MWA beam shapes \citep{Sokowlski2017}) and MWA observations (e.g. database of solar observations, calibration database \citep{Sokolwski2020}). 

All functions of these modules can broadly be divided into two major categories -- Calibration block and  Imaging block. Instead of describing these modules function-by-function, we present the workflows of these two major blocks in the Sections \ref{sec:calibration_block} and \ref{sec:image_block} respectively. Interested users can find the details of these functions in the documentation of P-AIRCARS available online.  
\begin{figure*}
    \centering
    \includegraphics[trim={1cm 7cm 1cm 1cm},clip,scale=0.55]{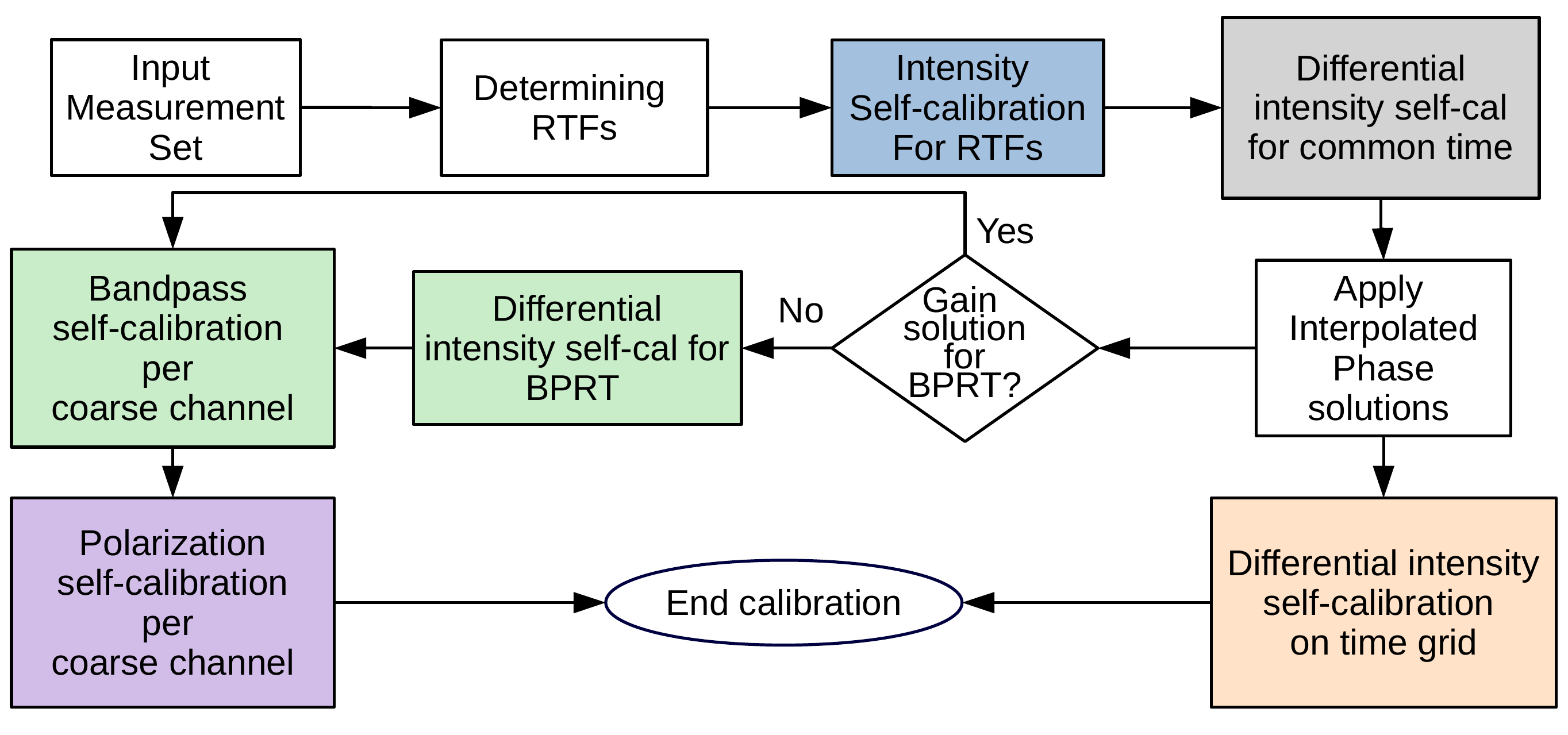}
    \caption{{\bf Flowchart describing the calibration block of P-AIRCARS.} Major stages of the calibration block are shown by color boxes. }
    \label{fig:calibration_block}
\end{figure*}

\begin{figure}
    \centering
    \includegraphics[trim={6.7cm 1.4cm 3cm 2cm},clip,scale=0.4]{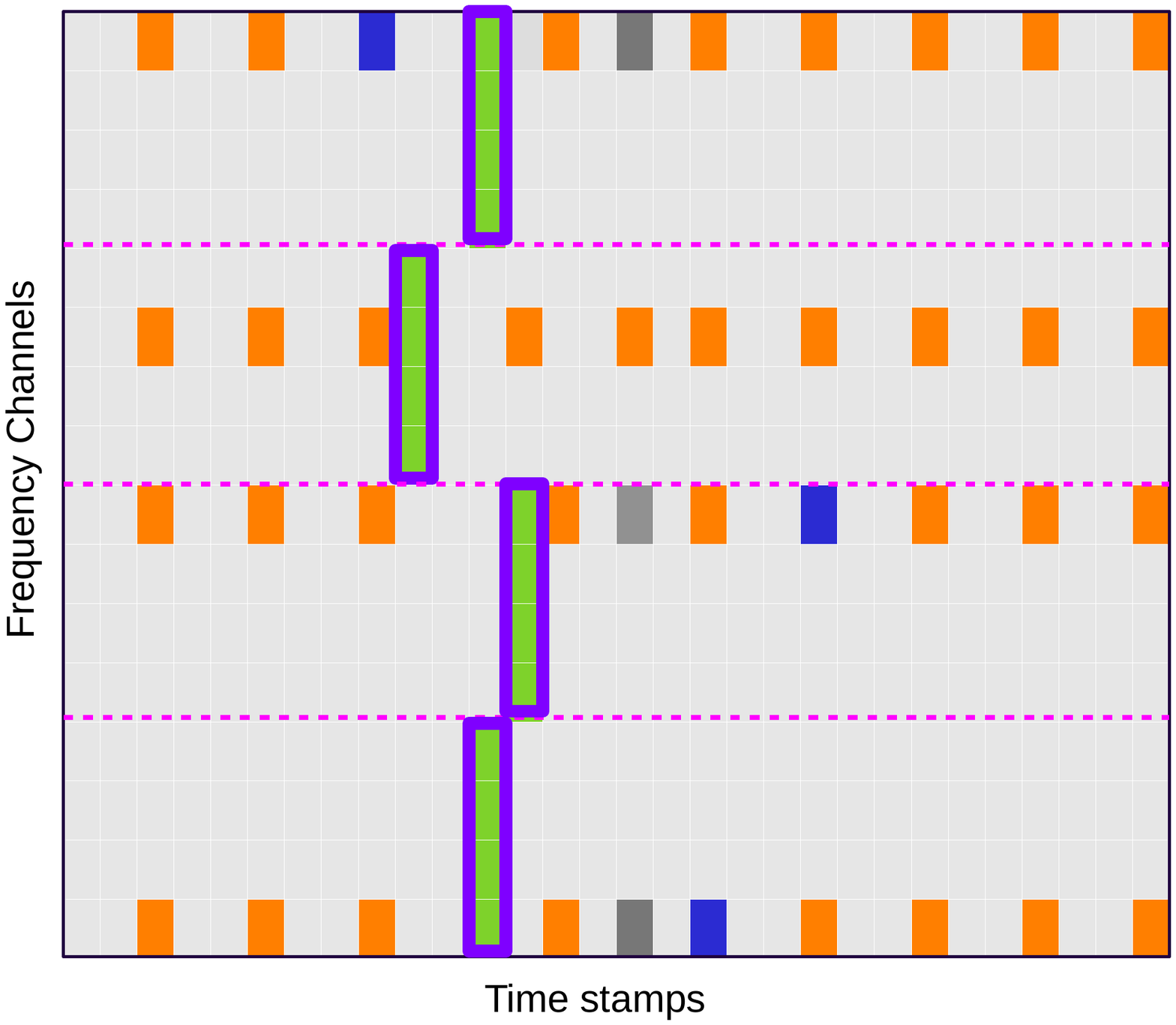}
    \caption{{\bf Time-frequency grid for parallel calibration.} Blue blocks represent the RTFs. Phase part of the gain solutions are interpolated on a common timeslice shown by the dark grey cells. Orange blocks represent the time and frequency slices where differential intensity self-calibration is performed. Bandpass and polarization calibrations are performed for individual coarse channels, which are marked by green. Pink dotted lines demarcate the 1.28 $\mathrm{MHz}$ wide coarse channels.}
    \label{fig:parallel_mechanism}
\end{figure}

\subsection{Implementation of Calibration Block} \label{sec:calibration_block}
The first major block of the P-AIRCARS is the calibration block. Calibration is done in three steps as discussed in Section \ref{sec:paircars_cal_algorithm}. 
Since the antenna gains vary over time and frequency, in principle, one should perform self-calibration for each timestamp and frequency channel independently. 
Due to the intrinsic spectro-temporal variability of solar emissions, one is forced to make an independent source model for every time and frequency slice during self-calibration, which makes self-calibration for every time and frequency slice extremely compute-intensive. As the calibration for each of the time and frequency slices are independent, it can be cast in an embarrassingly parallel framework, which is implemented in P-AIRCARS. The flowchart of the calibration block is shown in Figure \ref{fig:calibration_block}. 

To start the process of calibration, a maximum of three 1.28 MHz coarse channels are chosen
spanning the entire bandwidth of the data. 
Each of these chosen spectral channels is defined as a ``Reference frequency" (RF). Next, a time slice, defined as ``Reference time" (RT), is chosen for each of these RFs separately on which to perform the calibration. These are referred to as ``Reference Time and Frequency" (RTF) slices.

Figure \ref{fig:parallel_mechanism} shows an example with four coarse channels with their boundaries marked by dashed magenta lines. RTFs are shown by blue cells in this figure. 
If calibrator observations are available, P-AIRCARS first applies the gain solutions obtained from them. Otherwise, intensity self-calibration is initiated on the raw data. 
The calibration process is initiated using the highest time resolution available in the data
and if necessary the temporal span of the data used for calibration is progressively increased in an attempt to arrive at reliable gain solutions. Care is taken to not exceed the timespans over which solar emissions or ionospheric conditions are expected to evolve. The default value of this maximum timespan is set to 10$\mathrm{s}$.
Once this is done the pipeline moves to the next stage, namely bandpass self-calibration.

\iffalse
\begin{figure*}
    \centering
    \includegraphics[trim={5cm 4cm 4cm 1.5cm},clip,scale=0.8]{compare_DS_with_image.pdf}\\
    \includegraphics[trim={3cm 9cm 4cm 2cm},clip,scale=0.75]{quietSun_typeIII_collage.pdf}
    \caption{{\bf Comparison of the dynamic spectrum with the flux calibrated brightness temperature maps. Top panel: } Dynamic spectrum obtained following the method developed by \citet{oberoi2017}. {\bf Bottom panel:} Left image is for 06:12:50.25 UTC and the right image is for 06:13:06.25 UTC. For 06:12:50.25 UTC, marked by a green dot, the average $T_\mathrm{B}$ obtained from the dynamic spectrum is 0.92 $\mathrm{MK}$, which is close to the disc averaged value, 0.7 $\mathrm{MK}$, obtained from the image. For 06:13:06.25 UTC, marked by a cyan dot, the average $T_\mathrm{B}$ obtained from the dynamic spectrum is 18 $\mathrm{MK}$, which is also similar to the disc averaged value, 23 $\mathrm{MK}$, obtained from the image. The size of the disc is shown by the blue circles, which are 40 arcmins in size.}
    \label{fig:compare_ds_image}
\end{figure*}
\fi

\citet{Sokolwski2020} demonstrated that the variation of phases across the 80--300 MHz band for the MWA antenna tiles can be well modeled by a straight line, though the amplitudes show more complex variations.
Hence, it is reasonable to interpolate the phases across the MWA band using a linear model.
The phase variations over a large bandwidth cause a significant frequency-dependent shift of the source from the phase center. 
To avoid this problem, the phase part of the gain solutions are interpolated across the entire observing band, while the amplitudes are held constant at unity. Phases are interpolated across frequency at a common timeslice marked by grey cells in Figure \ref{fig:parallel_mechanism}. We define a time grid marked by orange cells in Figure \ref{fig:parallel_mechanism} for each of the coarse channels, and the differential gain solutions are computed in parallel. 
Simultaneously, bandpass and polarization calibrations
are performed at ``band-pol reference time (BPRT)" individually for each coarse channels marked by green boxes with purple borders in Figure \ref{fig:parallel_mechanism}.
If the gain solutions are not available at BPRT, a differential intensity self-calibration is performed at BPRT at RF.

Once all the calibrations are complete, this information is compiled in a single calibration table spanning the entire time and frequency range. Linearly interpolated gain solutions are drawn from this final calibration table and applied during imaging. The calibration block requires one to identify RTF and BPRT. The criteria for the choice of RTF and BPRT are discussed in Sections \ref{subsec:rtf_choice} and \ref{subsec:bprt_choice} respectively. Prior to this, P-AIRCARS identifies any bad data from the solar dynamic spectrum as described in Section \ref{subsec:bad_data_ds} and excludes them to avoid any problem during calibration.

\begin{figure*}[!htbp]
    \centering
    \includegraphics[trim={0.4cm 0cm 0.6cm 0cm},clip,scale=0.55]{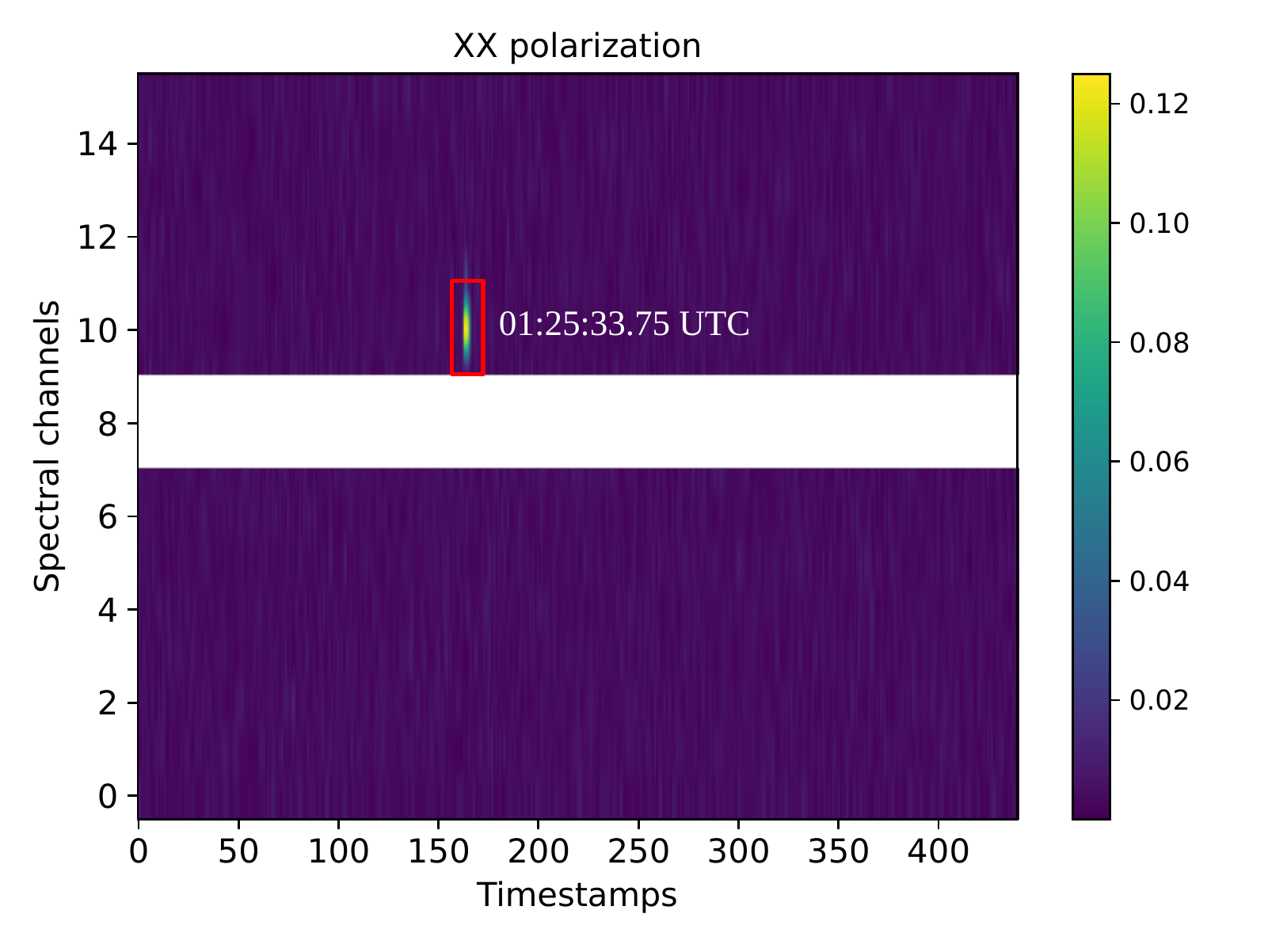}~\includegraphics[trim={0.4cm 0cm 0.5cm 0cm},clip,scale=0.55]{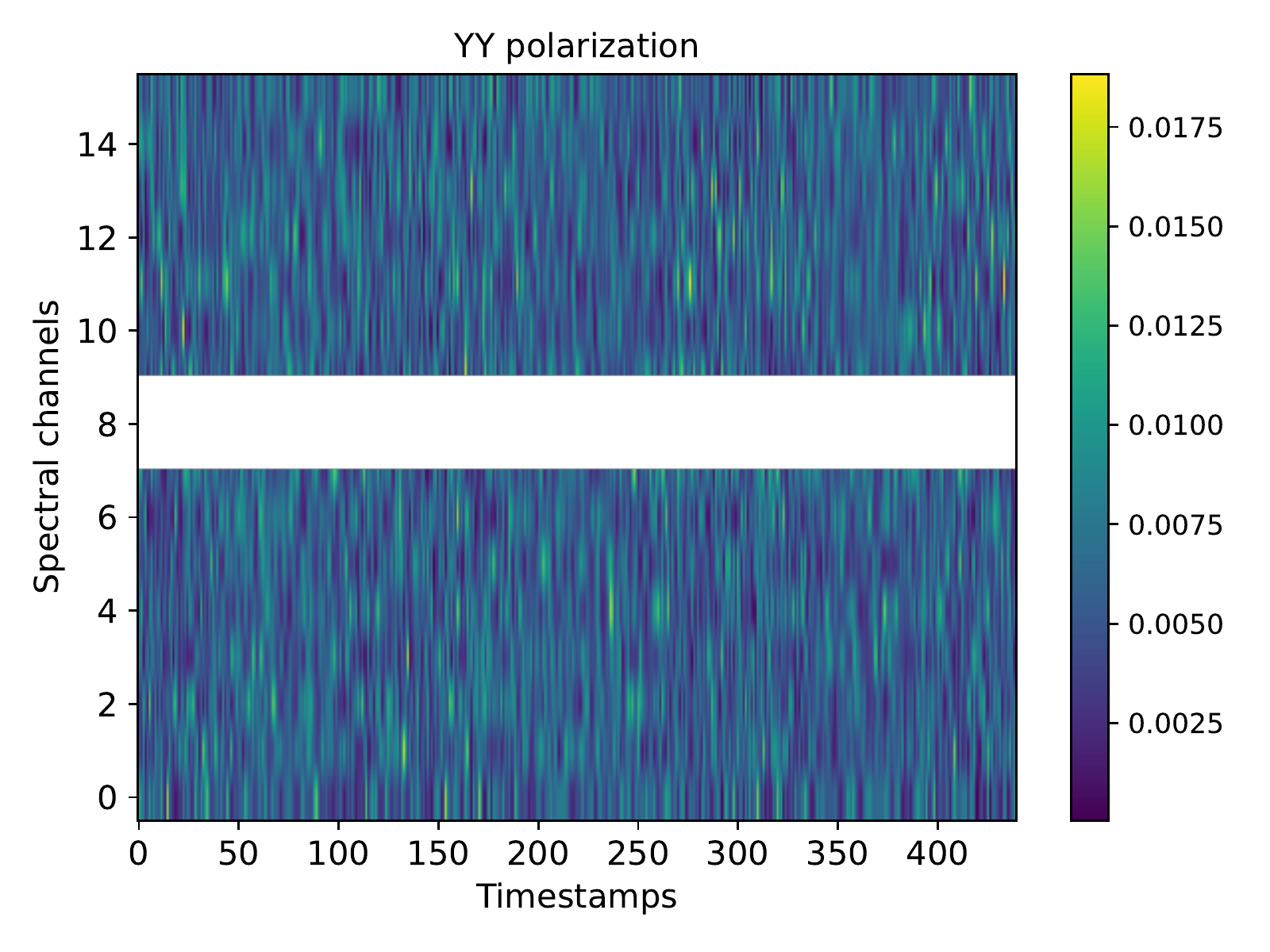}\\
     \includegraphics[trim={0.1cm 0cm 0cm 0.1cm},clip,scale=0.45]{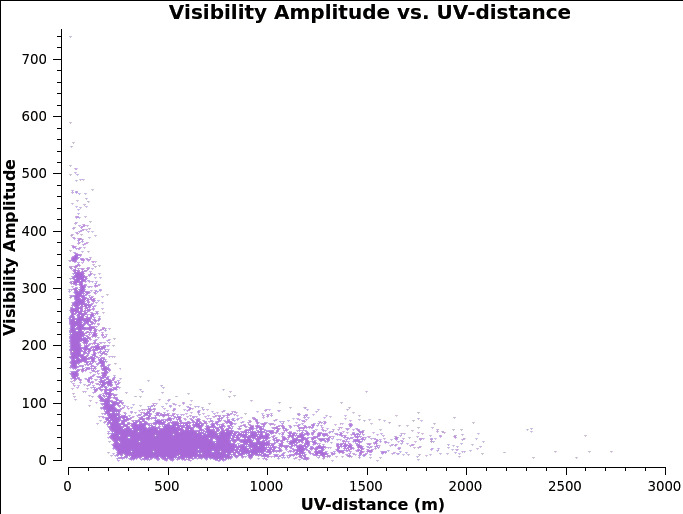}\includegraphics[trim={0.1cm 0cm 0cm 0.1cm},clip,scale=0.45]{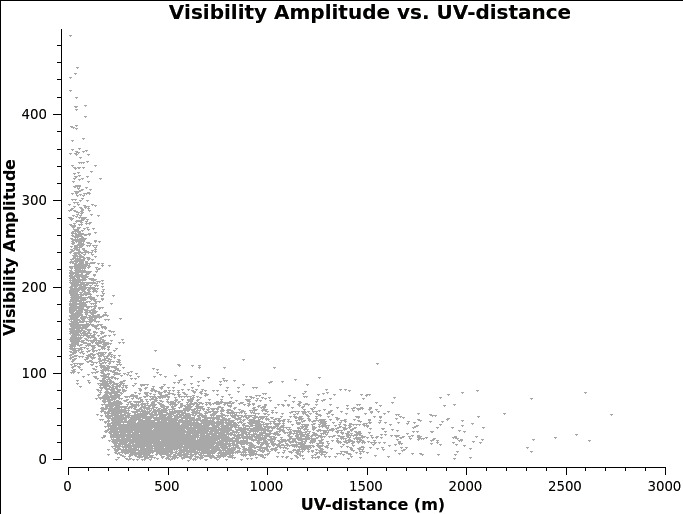}\\
     \vspace{0.5cm}
     \includegraphics[trim={0.1cm 0cm 0cm 0.1cm},clip,scale=0.45]{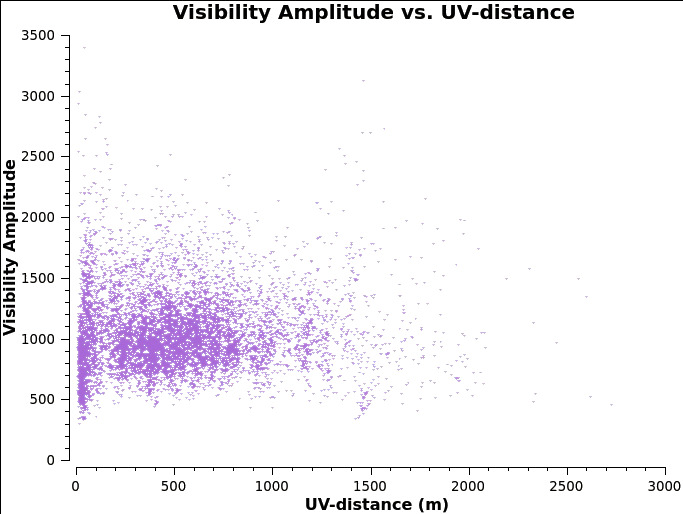}\includegraphics[trim={0.1cm 0cm 0cm 0.1cm},clip,scale=0.45]{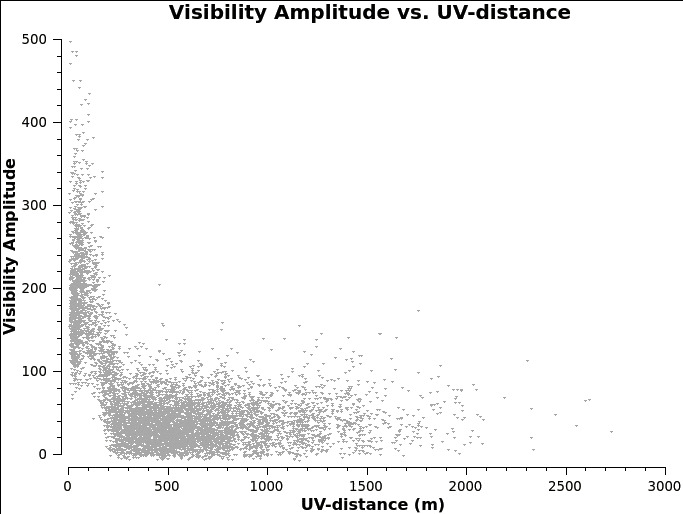}
    \caption{{\bf Demonstration of flagging of the bad data based on solar dynamic spectrum. Top panel :} It shows the dynamic spectrum of $r_\mathrm{a}$. %The left panel shows the dynamic spectrum for the XX polarization and the right panel shows it for the YY polarization. 
    {\bf Middle panel :} The visibility amplitude for a time without any RFI/instrumental issue is plotted against the {\it uv-}distance. {\bf Bottom panel :} The visibility amplitude of a timestamp, 01:25:33.75 UTC, affected by RFI/instrumental issues is plotted against the {\it uv-}distance. This time slice is marked by the red box in the top left panel. In all rows, the left and the right panels represent the XX and the YY polarizations respectively.}
    \label{fig:ds_flag}
\end{figure*}

\subsubsection{Identifying Bad Data from the Dynamic Spectrum for Solar Observations}\label{subsec:bad_data_ds}
Even though the MWA is situated in an exceptionally low radio frequency interference (RFI) environment and is a very stable instrument, occasionally the MWA data does suffer from RFI and/or instrumental issues. It is important to ensure that only healthy data  i.e. data unaffected by RFI and/or instrumental issues) are examined while determining the BPRT and RTF. Sometimes, active solar emissions can mimic bad data, making it hard to identify bad data based on statistical characteristics in the time and frequency plane alone. 

We use the fact that for the MWA the amplitude distribution with {\it uv-}distance for active/quiet Sun emissions and data affected by RFI/instrumental issues are remarkably different. For the quiet Sun, the visibility distribution represents a disc of about 40 arcmins. It has been found that the compact sources usually associated with active emissions are slightly resolved at MWA resolution \citep{Mohan2019a,Mohan2021a,Mondal2021a,Mohan2021b}. This implies that the visibility distribution for these slightly resolved sources must show a slow drop in amplitudes with increasing baseline length. On the other hand, the small footprint of the MWA and the fact that the RFI sources are mostly far away imply that the entire array tends to see the same RFI environment, and shows a relatively constant visibility amplitudes distribution with {\it uv-}distance.
Both solar emission and RFI can vary by multiple orders of magnitude with time and frequency. Hence, we device a quantity, $r_\mathrm{a}$, that is insensitive to the magnitude of the visibility amplitudes themselves but relies on their distribution as a function of baseline length to identify bad data. $r_\mathrm{a}$ is defined as the ratio of mean visibility amplitudes of long (longest 10~\%) and short baselines ($<20\ \mathrm{m}$).

Figure \ref{fig:ds_flag} shows an example to illustrate the efficacy of this approach. The top panels show the dynamic spectra of $r_\mathrm{a}$. The middle panels show the amplitude distribution for a time and frequency slice for XX (left panel) and YY (right panel) polarization without any RFI/instrumental issues. The bottom panels show the amplitude distribution for XX (left panel) and YY (right panel) polarization for a timeslice (01:25:33.75 UTC) with RFI/instrumental issues. The differences in the visibility distributions for these two panels are self-evident. The data for only the XX polarization are affected
and are identified with high contrast in the top left panel. 
This demonstrates the capability of $r_\mathrm{a}$ dynamic spectra to unambiguously and efficiently identify the data affected by RFI/instrumental issues.
Finer levels of identification and flagging of bad data are carried out during later stages of analysis (Section \ref{sec:ankflag}).

\subsubsection{Choice of Reference Time and Frequencies}\label{subsec:rtf_choice}
The calibration solutions from the RT are applied to all other timestamps as the initial gain solutions. Hence, it is important to choose a timeslice that enables us to determine gain solutions for each of the antenna tiles with good signal-to-noise. An additional requirement is that the image for this timeslice should also show the quiet Sun disc with sufficient fidelity so that it can be used for alignment of solar images as discussed in Section \ref{subsec:solar_phase_shift}. For the current levels of imaging fidelity achievable with P-AIRCARS using MWA data, these requirements are typically met when a compact source with $T_\mathrm{B}\leq\ 10^7\ \mathrm{K}$ is present on the Sun. 

\begin{figure}
    \centering
    \includegraphics[trim={0.3cm 0.3cm 0.5cm 0cm},clip,scale=0.55]{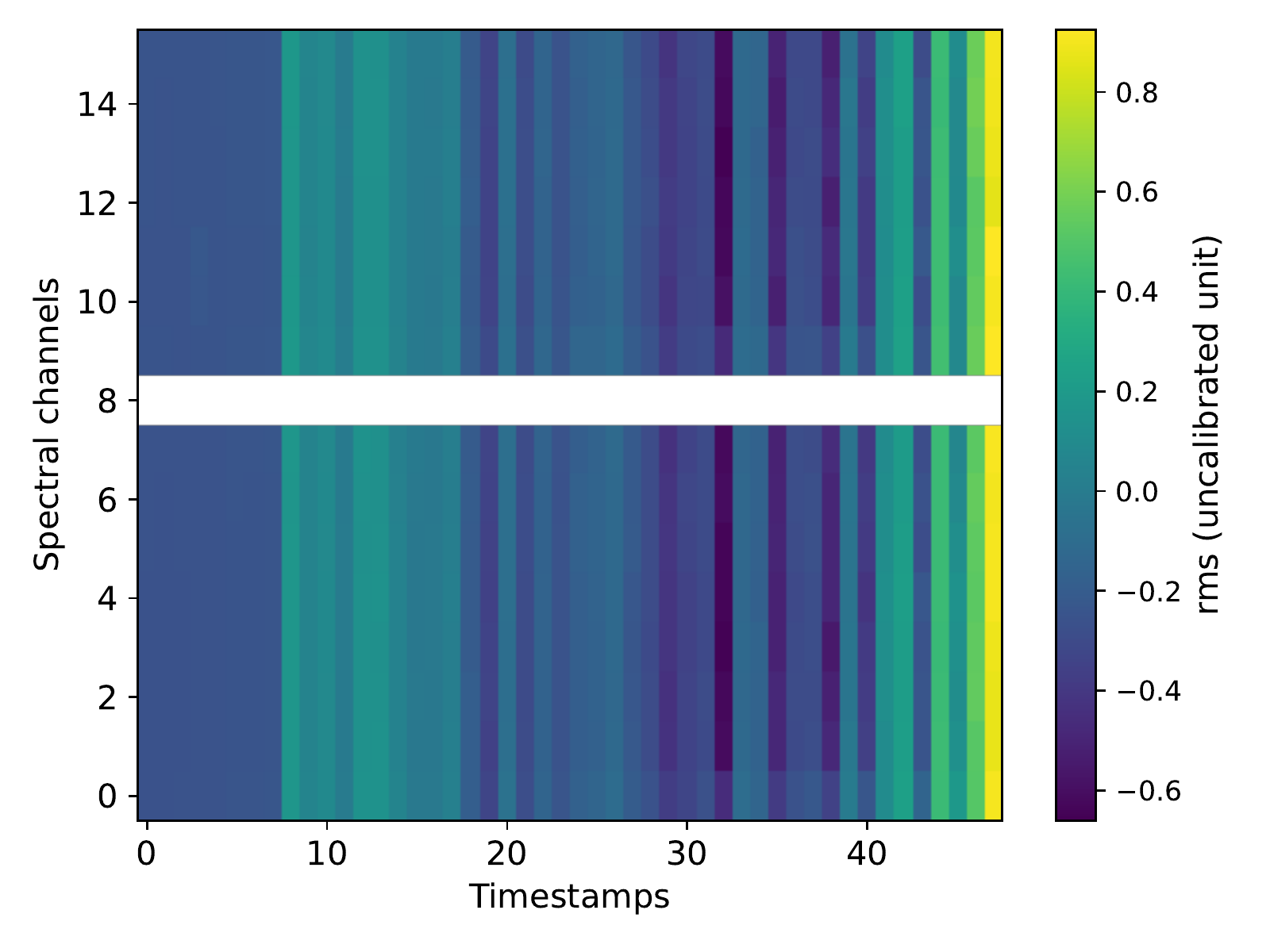}
    \caption{{\bf Dynamic spectrum of image rms.} The spectral and temporal span of the dynamic spectrum is 1.28 MHz and 240 s respectively. There is significant temporal variation in the image rms, while that along the spectral axis is barely evident.}
    \label{fig:rms_ds}
\end{figure}

The MWA is coherent enough to be able to proceed with imaging without any prior gain solutions \citep{Kansabanik_principle_AIRCARS}.  We have found that the rms noise of the dirty images can vary across time due to changes in solar flux density, but it does not vary drastically across frequency. An example dynamic spectrum of the rms measured far away from the Sun is shown in Figure \ref{fig:rms_ds} which illustrates these characteristics. We have examined several datasets and established that the temporal variations of the rms noise are largely independent of the spectral channel over this small bandwidth. At first, time slices that meet the $T_{\mathrm{B}}\leq\ 10^7\ \mathrm{K}$ requirement are identified using the flux density calibrated dynamic spectrum obtained using the method developed by \citet{oberoi2017}.
Dirty images are then made for every time slice meeting this requirement, for a single arbitrarily chosen spectral slice. The time slice with the highest imaging dynamic range is chosen to be the RT. The RF channel is identified next by following a similar procedure along the frequency axis for the chosen RT. 

\subsubsection{Choice of Band-pol Reference Time}\label{subsec:bprt_choice}
As the requirements for the bandpass and polarization calibration are different from those for the initial calibration, the criteria for the choice of BPRT are also different from those for the RTF. For reasons discussed in detail by \citet{Kansabanik2022,Kansabanik2022_paircarsI}, bandpass and polarization calibration require data taken under quiet solar conditions. The quiet solar time is identified in the given data, using the flux density calibrated dynamic spectrum arrived at following the method developed by \cite{oberoi2017}. 
The timestamps with $T_\mathrm{B}$ varying between $10^5-10^6\ \mathrm{K}$ are chosen to represent the quiet sun times. Among these, the timestamp with the maximum DR obtained from frequency averaged dirty images is selected as the BPRT. 

\subsubsection{Alignment of the Center of Solar Radio Disc}\label{subsec:solar_phase_shift}
\begin{figure*}[!htp]
    \centering
    \includegraphics[trim={1cm 12cm 5cm 0cm},clip,scale=0.75]{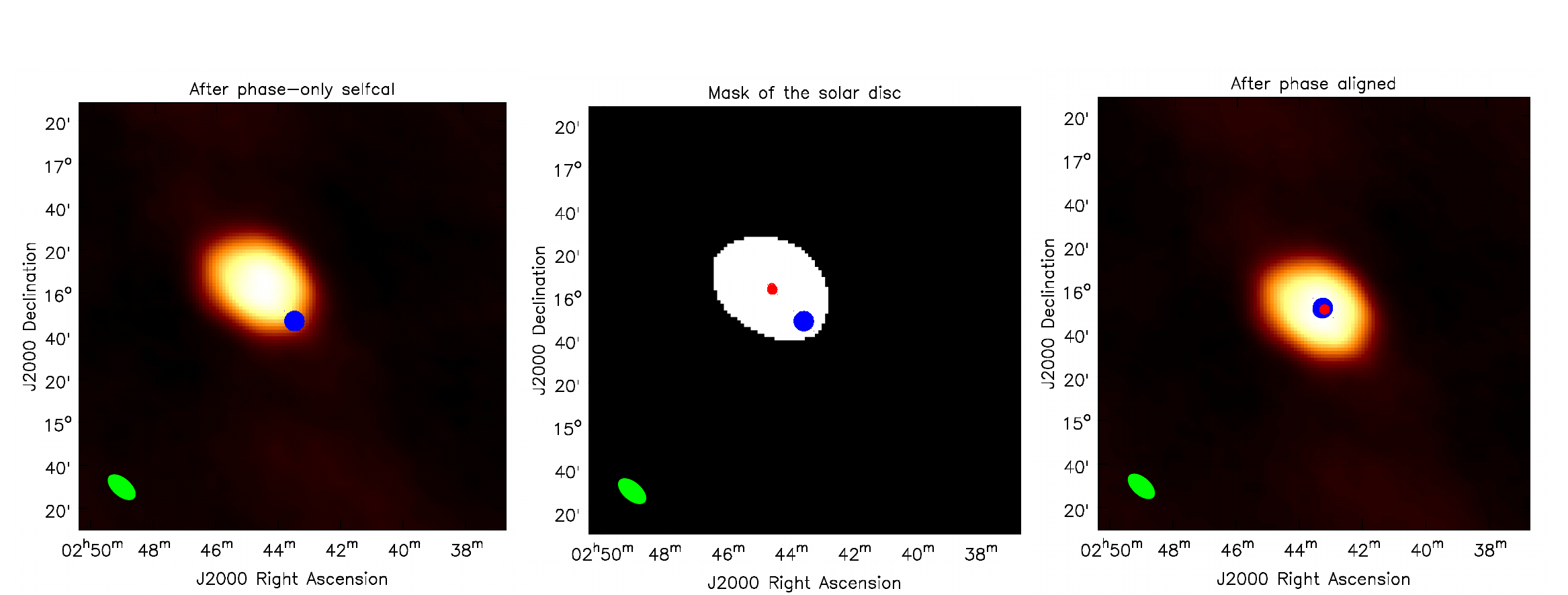}
    \caption{{\bf Alignment of the solar radio disc center with the optical solar disc center. Left panel: }Image after phase-only self-calibration. The center of the optical solar disc shown by the blue dot is not at the center of the radio disc. \textbf{Middle panel: } It shows the mask of the solar disc and the red dot represents the center of the radio disc. \textbf{Right panel: }Final image after alignment. The center of the optical and radio disc coincide after the alignment.}
    \label{fig:phase_align}
\end{figure*}

A common problem for any self-calibration-based approach is the loss of information about the absolute phase. Hence, the images for the RTFs are aligned using an image-plane-based method. P-AIRCARS first performs phase-only intensity self-calibration followed by amplitude-phase self-calibration \citep{Mondal2019,Kansabanik2022_paircarsI}.
Once the phase-only intensity self-calibration has converged, an image with the well-demarcated solar disc is available (Left panel of Figure \ref{fig:phase_align}). 
The blue circle marks the phase center of the radio image, which is set at the center of the optical disc. 

The region with more than $20\sigma$ detection significance is considered to be the solar disc, where $\sigma$ is the rms noise in the image measured close to the Sun. To avoid the intensity weighting, we define a mask with all the regions more than $20\sigma$ set to unity and the rest of the image set to zero as shown in the middle panel of Figure \ref{fig:phase_align}. The center of mass of the masked region is chosen to be the center of the solar radio disc marked by the red circle. The phase center of the source model is shifted to align with the blue circle. Using this aligned source model, a few rounds of phase-only self-calibration are performed.  
The final set of self-calibration solutions is then applied to the entire dataset to bring it to a common phase center.

\subsubsection{Flux Density Calibration}\label{subsec:flux_calib}
Another common limitation of any self-calibration-based approach is the loss of information about the absolute flux density scale. At the MWA, when dedicated calibrator observations are available with the same spectral and attenuation configuration as solar observation, an absolute flux density scale is obtained from the gain solution of the calibrator observations. When no calibrator observation is available with the above-mentioned criteria, P-AIRCARS does flux density calibration using an independent method developed by \citet{Kansabanik2022}.

\subsection{Implementation of Imaging Block}\label{sec:image_block}
\begin{figure*}[!htp]
    \centering
    \includegraphics[trim={1cm 1cm 1cm 1cm},clip,scale=0.55]{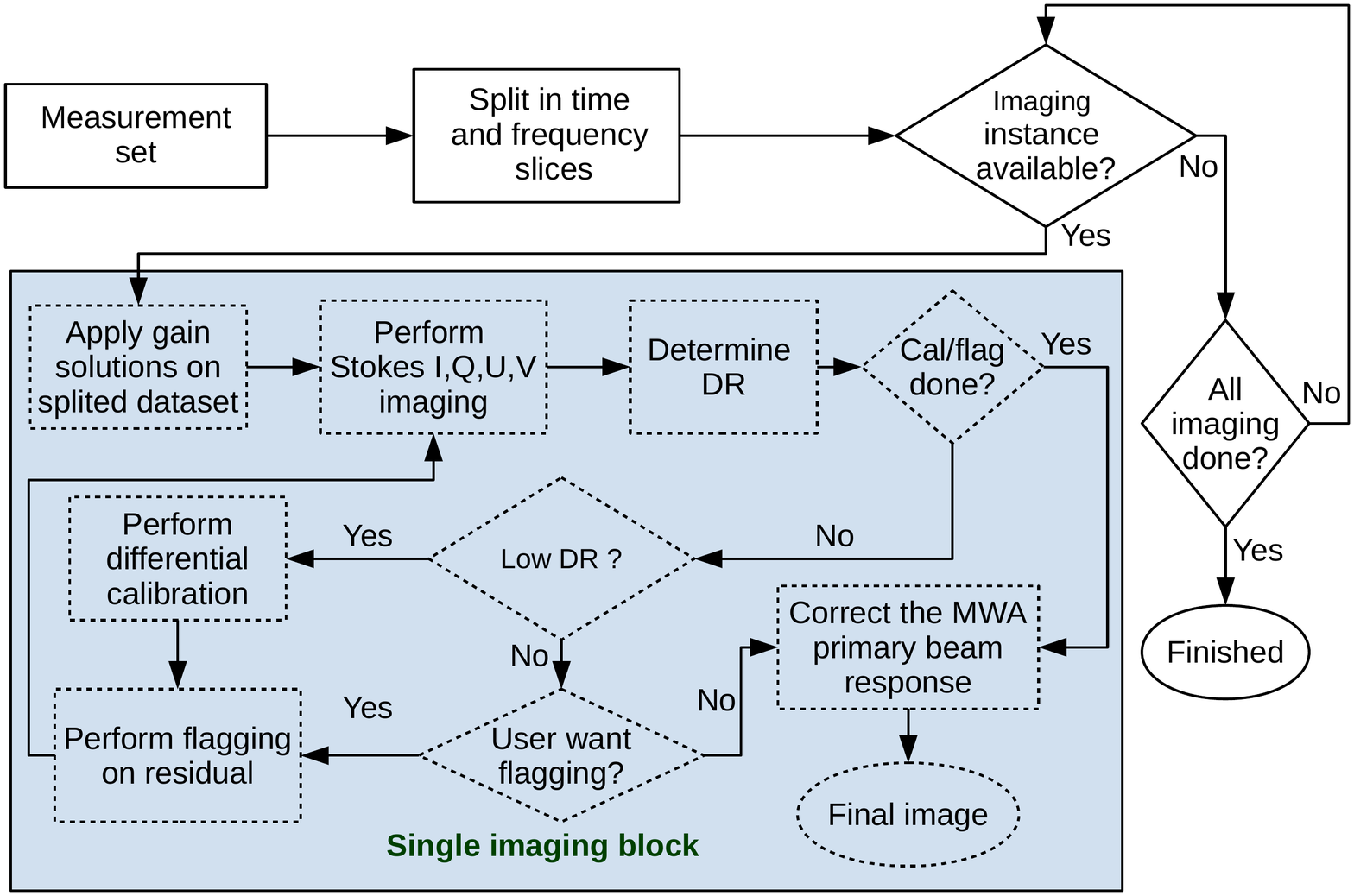}
    \caption{{\bf Flowchart describing the imaging block of P-AIRCARS.} A single imaging block is shown inside the blue shaded box.}
    \label{fig:imaging_block}
\end{figure*}

Once calibration solutions spanning the time and frequency ranges of interest are available, P-AIRCARS proceeds to image. In addition to imaging, this block also corrects the images for the instrumental primary beam. The problem is essentially embarrassingly parallel, and hence straightforward. The key requirement here is to allow the user to allocate a chosen fraction of the compute resources to P-AIRCARS and for P-AIRCARS to make the optimal use of these resources. This is achieved using a custom-developed parallelization algorithm described in Section \ref{subsec:parallel_imaging}. The flowchart of the entire imaging block is shown in Figure \ref{fig:imaging_block}. The functionality in the blue box marked as `single imaging block' is executed in parallel for the different time and frequency slices and is described in Section \ref{subsec:single_imaging}. 

\subsubsection{Parallelization of Imaging Block}\label{subsec:parallel_imaging}
As mentioned in Section \ref{sec:key_principles}, the total number of images to be produced can be as many as $370,000$ for observation with a 30.72 MHz bandwidth and 4 minutes duration.  The number of imaging threads required for this task is much larger than the compute capacity available with most machines. Hence, a scalable mechanism for their efficient parallelization is required. Whenever, the number of imaging jobs, $N_\mathrm{job}$, is smaller than the available CPU threads, $N_\mathrm{thread}$, all jobs are spawned simultaneously. Each job is assigned $n$ numbers of CPU threads, where $n$ is the integer closest to $N_\mathrm{thread}/N_\mathrm{job}$. Otherwise, P-AIRCARS allocates three CPU threads for each single imaging block. The $N_\mathrm{job}$, which can be spawned simultaneously, is given by:
\begin{equation}
    N_\mathrm{job}=\frac{N_\mathrm{thread}}{3}.
\end{equation}
Different imaging jobs may take different run times. To utilize the hardware resources efficiently, as soon as one imaging job is done, a new one is spawned. This process continues until all imaging jobs have been spawned.  

\subsubsection{Single Imaging Block}\label{subsec:single_imaging}
The single imaging block makes the image of a single time and frequency slice and is marked by the blue shaded box in Figure \ref{fig:imaging_block}. 
Imaging parameters are determined from the data, as discussed in Section \ref{subsec:imaging_params}. Users can choose to either do full polarimetric imaging or only total intensity imaging. 

First, the final calibration solutions are applied to the data. This is followed by a shallow deconvolution (10-$\sigma$ threshold) to ensure that no spurious emission gets included in the source model. Despite the shallowness of this deconvolution, it is sufficient to provide a good check for imaging quality. The DR of these images is compared with the minimum DR ($\mathrm{DR_{min}}$) of the images made during the process of calibration. If the DR of an image after shallow deconvolution is found to be smaller than a pre-defined fraction of $\mathrm{DR_{min}}$, an additional round of calibration is performed to account for the differential antenna gain variations which might have led to the drop in the DR. This pre-defined value is set to $10$\% by default. If the user chooses to perform flagging during the final imaging, independent of whether additional calibration is required or not, a single round of flagging is done on the residual visibilities using a custom-developed flagging software, \textsf{ankflag}, which is discussed in Section \ref{sec:ankflag}. Once the flagging is done, a single round of deep deconvolution is performed. These images are then corrected for the instrumental primary beam to arrive at the final images.

\section{Flagger for P-AIRCARS}\label{sec:ankflag}
Radio Frequency Interference (RFI) is the name given to the unwanted 
radio signals of human origin which are incident on 
a radio telescope and contaminate the cosmic signals. 
It is important to remove this contaminating signal from the data before calibration and imaging. This process is usually referred to as flagging and often the quality of the final images depends upon the efficacy of the flagging approaches and algorithms. Therefore, an automated calibration and imaging pipeline for high dynamic range imaging must include efficient and effective ways to identify and flag RFI.  
This is complicated by the fact that bright solar emissions can also mimic such RFI features in the time-frequency plane. In our experience, it is possible to improve upon the flagging performance offered by the in-built automated RFI flagging tasks in \textsf{CASA} \citep{mcmullin2007}.
To achieve this improvement, an independent flagging module, \textsf{ankflag}, has been included in P-AIRCARS.

Originally, \textsf{ankflag} was developed for an HI 21cm survey \citep{bera2019} with the upgraded Giant Metrewave Radio Telescope \citep[uGMRT;][]{Gupta_2017}, and has been used for several previous studies \citep[e.g.,][etc.]{Das_2020a,Das_2020,mondal2020_ptf10hgi,Das2022}. It has been found to be very efficient for flagging low-level RFI for the MWA solar observations and hence has been integrated in P-AIRCARS. 
\textsf{ankflag} is available as an independent module in P-AIRCARS. It can be used for RFI flagging of interferometric data from any radio interferometric array. 

\subsection{Basic Algorithm of \textsf{ankflag}}\label{subsec:ankflag_algorithm}
\textsf{ankflag} identifies bad or RFI-affected data as outliers to the Gaussian statistics which is assumed to represent the input data. Given this assumption, \textsf{ankflag} works best on model subtracted residual visibilities, and separately on the real and imaginary parts of the visibilities, as the amplitudes of the visibilities do not follow Gaussian statistics. Some of the basic algorithms for calculating the statistics of the input data are taken from  \textsf{FLAGCAL} \citep{Flagcal2012}, previously written for flagging and calibration of the interferometric data from the Giant Metrewave Radio Telescope \citep[GMRT;][]{Swarup_1991}.

The threshold for the outlier expressed as $X\sigma$, is set such that the expected value of elements outside the interval $[\mu-X\sigma,\ \mu+X\sigma]$ is less than 1, where, $\sigma$ is the rms and $\mu$, the mean of the data. A user-defined tolerance factor, $f$, is multiplied with this threshold so that the effective allowed range for the data is $[\mu-fX\sigma,\ \mu+fX\sigma]$. Elements outside this range are considered outliers and flagged. $f$ essentially accounts for the fact that the statistics of the data generally deviate from exact Gaussian statistics, and hence lead to over-flagging data while trying to flag very low-level RFI. We have found that a value of $f$ ranging from 1.7 to 1.9 works well for the MWA solar observations.

For outlier detection, user can use both mean$-$rms  or median$-$MAD statistics.
The mean$-$rms statistics can be easily biased by the outliers present in the data, while the median$-$MAD statistics are far more robust against the outliers. 
By default, P-AIRCARS configures \textsf{ankflag} to use the median$-$MAD statistics.

\textsf{ankflag} has two modes of operation $-$ \textsf{baseline} and \textsf{uvbin} mode. Both of these modes are discussed below :
\begin{enumerate}
    \item \textbf{baseline mode: }In the baseline mode, 
    in the first step, statistics are calculated from the time-frequency plane of every scan and baseline separately. Outlier visibilities are identified and flagged based on these statistics. In the next step, depending upon the choice of the user, statistics for all scans and baselines are compared with one another. 
    Since solar emission can show drastic variation in flux density over small durations and bandwidths, this mode is not suitable for solar observation but has been used successfully for other astronomical observations.
    
    \item \textbf{uvbin mode: }In the \textsf{uvbin} mode, all visibilities are binned in a user-defined number of two-dimensional bins in the Fourier plane of the sky brightness distribution ({\it uv}-plane), such that each bin has approximately the same number of visibilities. Visibilities in each bin are inspected separately to identify and flag outliers. The \textsf{uvbin} mode is usually slower than the baseline mode and relies on a good {\it uv-}coverage. Since the MWA has a very good {\it uv-}coverage, this mode is ideal for solar observations with the MWA and is used in P-AIRCARS. 
\end{enumerate}
In both \textsf{baseline} and \textsf{uvbin} mode, each input polarization is treated independently. \textsf{ankflag} is primarily written in C, with a ``wrapper” written in {\em python} to use it as an independent python module. \textsf{ankflag} uses the GNU scientific library \citep[GSL,][]{Galassi2018} for performing algebraic and statistical tasks. Details of the tasks are available in the documentation available online. 
\begin{figure}
    \centering
    \includegraphics[trim={0.1cm 0cm 0cm 0.1cm},clip,scale=0.44
    ]{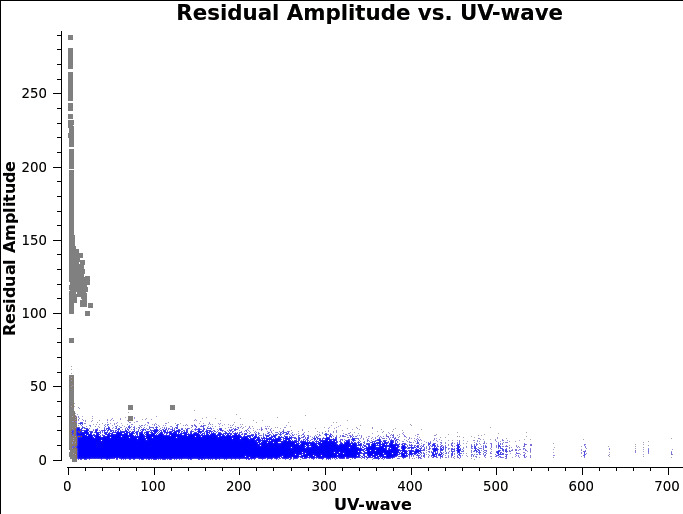}
    \caption{{\bf Demonstration of ankflag on the MWA data.} Flagging is done on a single time-frequency slice using {\it uv}-bin mode. Grey squares and blue points show the flagged and un-flagged data respectively.}
    \label{fig:ankflag_demo}
\end{figure}

\subsection{Demonstration of \textsf{ankflag}}\label{subsec:ankflag_demo}
For this demonstration, we use \textsf{ankflag} in \textsf{uvbin} mode on some example MWA solar data. The amplitudes of the residual visibilities against {\it uv-}wavelength are shown in Figure \ref{fig:ankflag_demo}. 
%The left and right panels show the data before and after performing the flagging respectively. 
The grey squares show the bad data, which are flagged. A very small amount of good data has been flagged. This demonstrates the capability of \textsf{ankflag} to selectively remove low-level RFI/bad data without over-flagging the good data.

\begin{table*}
\centering
\renewcommand{\arraystretch}{1}
    \begin{tabular}{|p{0.4cm}|p{0.4cm}|p{1cm}|p{1.0cm}|p{1.5cm}|p{1.5cm}|p{1.2cm}|p{2cm}|p{1.7cm}|p{1.6cm}|p{1.6cm}|}
    \hline
       QF & RF & $\mathrm{th_{start}}$ & $\mathrm{th_{step}}$ & $\mathrm{th_{stop}}$ & $g_\mathrm{min, SNR}$ & $\Delta$ DR &  $t_\mathrm{interval}$ (s) & $\mathrm{DR_{max}}$ & $f_\mathrm{{res}}$\\ \hline \hline 
       0 & 0 & 9.0 & 1.0 & 6.0 & 2.5 & 25 & 30 & 100 & 0.03\\
       \hline
        0 & 1 & 9.0 & 1.0 & 6.5 & 3.0 & 22 & 20 & 500 & 0.03\\
       \hline
         0 & 2 & 9.0 & 1.0 & 7.9 & 3.5 & 20 & 15 & 1000 & 0.03\\
       \hline
        1 & 0 & 10.0 & 0.5 & 6.0 & 3.5 & 20 & 15 & 1000 & 0.015\\
       \hline
        1 & 1 & 10.0 & 0.5 & 6.5 & 4.0 & 18 & 10 & 5000 & 0.015\\
       \hline
        1 & 2 & 10.0 & 0.5 & 7.0 & 4.0 & 15 & 7 & 10000 & 0.015\\
       \hline
        2 & 0 & 11.0 & 0.25 & 6.5 & 4.0 & 18 & 10 & 10000 & 0.01\\
       \hline
        2 & 1 & 11.0 & 0.25 & 7.0 & 4.5 & 15 & 7 & 50000 & 0.01\\
       \hline
        2 & 2 & 11.0 & 0.25 & 7.0 & 4.5 & 12 & 5 & 100000 & 0.01\\
       \hline
    \end{tabular}
    \caption{Self-calibration parameters for different combinations of QF and RF.}
    \label{table:cal_params}
\end{table*}

\section{Calibration and Imaging Parameters}\label{sec:determine_params}
For reasons discussed in Section \ref{sec:key_principles}, P-AIRCARS is designed to determine the parameters for calibration and imaging in an unsupervised manner. There are only two high-level parameters that the user needs to specify to guide the choices to be made by P-AIRCARS.
These are \textsf{quality\_factor} (QF) and \textsf{robustness\_factor} (RF). Both of these parameters take three integer values : 0, 1, and 2. QF relates to the choices of parameters impacting the final image quality, with a higher number corresponding to a better imaging quality. Similarly, RF relates to choices made regarding the convergence criteria and robustness of the self-calibration. The final choice of calibration parameters depends upon the combination of QF and RF chosen, though the final imaging parameters depend only on the choice of QF. In general, larger numbers for QF and RF lead to larger computational loads and hence longer run times.

\subsection{Calibration Parameters}\label{subsec:calib_params}
Multiple different parameters need to be specified for calibration tasks. These include the solution interval along the temporal axis ($t_\mathrm{interval}$), the minimum acceptable signal-to-noise of the antenna gain solutions ($g_\mathrm{min, SNR}$), the shortest baselines to be used, and the changes in DR ($\Delta$ DR) over the past few images defining the convergence of the self-calibration process. The length of the shortest baseline is chosen to avoid contributions from the Galactic diffuse emission as it is hard to model and can dominate the solar signal. By default, P-AIRCARS excludes visibilities below $3\lambda$, which corresponds to $\sim$20$\ \mathrm{degree}$ in angular scale. 

Some additional parameters also need to be specified for the self-calibration process. During intensity self-calibration, deconvolution thresholds are decreased in steps with the self-calibration iterations. The start, stop and increment values for these thresholds, $th_{start}$, $th_{stop}$ and $th_{step}$ respectively. These are specified in units of image rms measured far away from the Sun, $\sigma$. We define another quantity, the fractional residual flux density, which is the ratio of disc integrated flux densities obtained from the residual and solar images from the latest self-calibration iteration. Starting from $\mathrm{th_{start}}$ the deconvolution threshold is lowered by $\mathrm{th_{step}}$ until it either reaches $\mathrm{th_{stop}}$ or the fractional residual flux density, $f_\mathrm{{res}}$, drops below some pre-defined thresholds listed in Table \ref{table:cal_params}.  If the imaging DR exceeds a pre-defined threshold, $\mathrm{DR_{max}}$, the self-calibration process is stopped even though it might not have converged. 
The numerical values of all of these parameters chosen based on the combination of QF and RF are listed in Table \ref{table:cal_params}.

\subsection{Imaging Parameters}\label{subsec:imaging_params}
Multiple different parameters need to be specified for imaging. These include -- the size of the image, pixel size, the {\it uv-}taper parameter, visibility weighting scheme, the choice of scales for multiscale deconvolution, the deconvolution threshold, the deconvolution gain, and whether or not to use {\it w-}projection. P-AIRCARS is designed to provide default values of each of these parameters. The expert user always has the flexibility to override the defaults. The default values for the image pixel size and {\it uv-}taper value are estimated from the data.
The default values of image size, deconvolution threshold, deconvolution gain and use of {\it w-}projection are decided based on the choice of the QF. The default values of visibility weighting used during imaging and Gaussian scales used for multiscale deconvolution are chosen independent of the data and QF values. Details of how the default values of imaging parameters are arrived at are available in the documentation available online.  

\section{P-AIRCARS Features}\label{sec:features}
This section briefly highlights some salient features of P-AIRCARS. 
\begin{enumerate}
\item \textbf{Modularity:} As described in Section \ref{sec:pipeline_arch}, P-AIRCARS architecture is highly modular. This not only makes it easy to maintain and upgrade, but it also enables P-AIRCARS to offer the possibility of using multiple different radio interferometric packages. 

\item \textbf{Ease of use:} To facilitate the use by community members with little or no prior experience in radio interferometry, P-AIRCARS provides reasonable defaults for all parameters, which can be overwritten by experience users. 

\item \textbf{Input validation:} For P-AIRCARS to run successfully, all of the inputs need to be consistent and compatible with the data. To ensure this, P-AIRCARS first checks for this consistency and compatibility before initiating processing. In case some inconsistent or incompatible inputs are found, their values are reset to the default values and a warning is issued to the user.

\item \textbf{Fault-tolerant:} To be able to deal with a wide variety of solar and instrumental conditions in an unsupervised manner, P-AIRCARS has been designed to be fault-tolerant. When it fails, it tries to make data-driven decisions about updating the relevant parameter values to overcome the source of the problem.  

\item \textbf{Notification over e-mail:} Typical run-time for P-AIRCARS for MWA data can run into days. To make it convenient for the users to stay abreast of its progress, P-AIRCARS can provide regular notifications about its status to a user-specified list of e-mail addresses.

\item \textbf{Graphical User Interface:} P-AIRCARS provides a Graphical User Interface (GUI) for specifying values of input parameters. P-AIRCARS saves a detailed log of the various processing steps and also provides a graphical interface to easily view it.
\end{enumerate}

\section{P-AIRCARS Requirements and Performance}\label{sec:requirement}
This section summarises the hardware and software requirements for P-AIRCARS and provides some information about its run-time for typical MWA data.

\subsection{Hardware Requirements}\label{sec:hardware}
P-AIRCARS is designed to be used on a wide variety of hardware architectures, all the way from laptops and workstations to HPCs. It uses a custom-designed parallelization framework, which also does the scheduling for non-HPC environments. P-AIRCARS has been tested with a minimum configuration of 8 CPU threads and 8 GB RAM, which is increasingly commonplace in commodity laptops. P-AIRCARS has also been tested on workstations with 40$-$70 CPU threads and 256 GB of RAM.  

\subsection{Software Requirements}\label{sec:software}
P-AIRCARS uses multiple radio interferometric software packages (e.g., \textsf{CASA}, \textsf{WSClean}, \textsf{CubiCal/QuartiCal}), each of which have multiple specific software dependencies. P-AIRCARS has been tested successfully on Ubuntu (20.04) and CentOS (7 and 8) Linux environments. P-AIRCARS requires Python 3.7 or higher. To reduce the tedium of dealing with dependency conflicts and make P-AIRCARS deployable out-of-the-box, we plan to containerize it using {\em Docker} \citep{docker2014}. While P-AIRCARS is under constant development, interested users can download the stable version described here from Zenodo \citep{paircars_zenodo}\footnote{\url{https://zenodo.org/record/7382624}}. 

\subsection{Assessment of Run-time}\label{sec:runtime}
\begin{figure}
    \centering
    \includegraphics[trim={0.44cm 0.5cm 0.3cm 0cm},clip,scale=0.54]{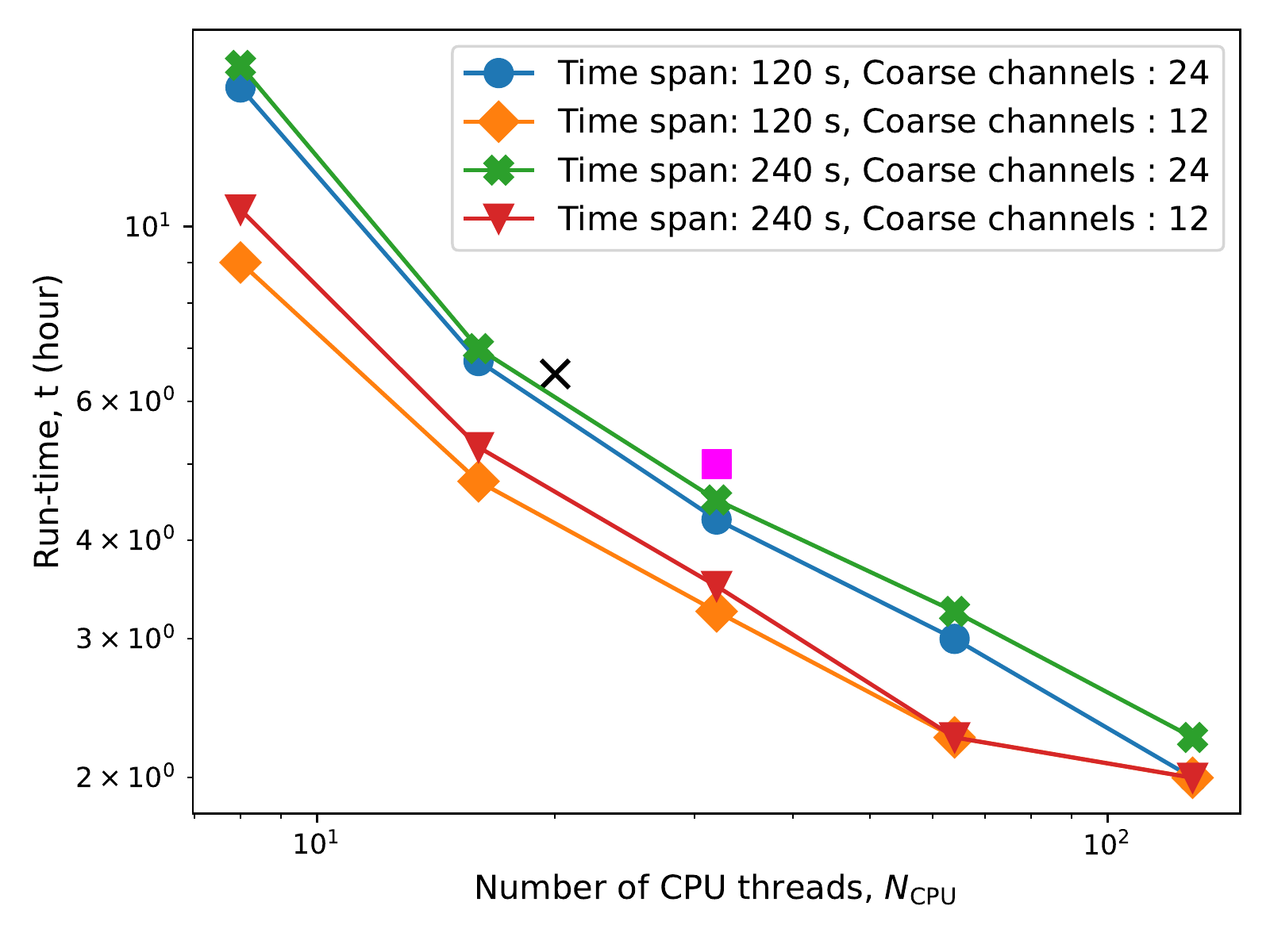}
    \caption{{\bf Variation of calibration time with the available number of CPU threads.} The green, orange, blue, and red points represent the expected run-time for a combination of temporal and spectral spans. The black cross and magenta square shows the run-time from a real dataset with 20 and 30 CPU threads respectively.}
    \label{fig:core_hours}
\end{figure}

To provide an overall estimate for P-AIRCARS run-time, we list the run times for individual processing blocks:
\begin{enumerate}
    \item Each RTF takes about an hour (marked by blue cells in Figure \ref{fig:parallel_mechanism}).
    \item Bandpass calibration for each coarse channel takes about 15 minutes (marked by green cells with purple borders in Figure \ref{fig:parallel_mechanism}).
    \item Polarization calibration for each coarse channel takes about 45 minutes (marked by green cells with purple borders in Figure \ref{fig:parallel_mechanism}).
    \item Each differential intensity self-calibration takes about 10 to 15 minutes (marked by dark grey and orange cells in Figure \ref{fig:parallel_mechanism}).
\end{enumerate}
The first three steps are done sequentially and add up to a minimum total run-time of about 2 hours. Differential intensity self-calibrations are all done in an embarrassingly parallel manner.

Figure \ref{fig:core_hours} shows the expected variation in run-time, $t$, taken for calibration as a function of the number of CPU threads, $N_\mathrm{CPU}$, for a few different combinations of temporal and spectral spans on a log scale. Orange and red points show the run-time for a dataset with 12 coarse channels with an observing duration of 120 and 240 seconds, respectively. The difference between the two curves is small at the low $N_\mathrm{CPU}$ end and grows even smaller with increasing $N_\mathrm{CPU}$.
At the large $N_\mathrm{CPU}$ end, when there are enough resources available to spawn all of the differential calibration jobs in parallel, there remains no difference in the corresponding $t$s. The blue and green points show the variation of $t$ with $N_\mathrm{CPU}$ for datasets with 24 coarse spectral channels for observing durations of 120 and 240 seconds, respectively and show similar behavior. Naturally, at the low $N_\mathrm{CPU}$ end, they take significantly longer than the 12 coarse channel datasets, and the difference between $t$ for datasets with 24 and 12 coarse channels reduces with increasing $N_\mathrm{CPU}$. These curves have been obtained using a model for P-AIRCARS performance. This model has been benchmarked using measured $t$ for a dataset with 24 coarse channels spanning 240 seconds and processed using 20 and 32 CPU threads, respectively, shown by a black cross and a pink square and lie close in Figure \ref{fig:core_hours}. These values are close to the predicted model values. 

Unlike calibration, imaging jobs are embarrassingly parallel with $t$ decreasing linearly with increasing $N_\mathrm{CPU}$. A MWA solar observation, with 30.72 $\mathrm{MHz}$ bandwidth and 4 minutes duration, leads to about $50,000$ images at 160 $\mathrm{kHz}$ and 0.5 $\mathrm{s}$ resolution. For such a dataset, P-AIRCARS typically requires about 4 hours for calibration and about 250 hours ($\sim10$ days) for imaging using 32 CPU threads. 

\section{Future Plans}\label{sec:limitation_and_future}
The modular design of P-AIRCARS allows it to benefit easily from the developments and improvements being continually made in the underlying software packages it uses. We plan to incorporate the recent developments of these software packages and data structures in P-AIRCARS. The next generation of {\it measurement set} format (MS-v.3) has recently been released\footnote{\url{https://casacore.github.io/casacore-notes/264.pdf}}, as a part of the Next Generation CASA infrastructure \textsf{(ngCASA)}\footnote{\url{https://cngi-prototype.readthedocs.io}} effort. The MS-v.3 offers a major advantage by significantly reducing the input-output (IO) overheads incurred during calibration and imaging processes. As IO forms a notable fraction of P-AIRCARS run-time, we expect this to bring significant benefits. Incorporating MS-v.3 in P-AIRCARS, however, needs calibration and imaging software to be compatible with the MS-v.3 data structures. This requirement is already met by \textsf{QuartiCal}, which P-AIRCARS relies upon for calibration.
 
While they have not been activated yet, P-AIRCARS has internal mechanisms for each run of P-AIRCARS to contribute calibration solutions to a global database. This has been done with a vision to build a central repository of all available calibration solutions for MWA solar data accessible to all P-AIRCARS users. It will benefit the individual users by providing them with pre-existing calibration solutions when available and reducing their run-time. Over time, as the usage of P-AIRCARS grows, we expect this to become a useful resource for the community.

While the P-AIRCARS architecture is compatible with HPC deployment, it has not been deployed on one yet, primarily due to a lack of a suitable opportunity.  Currently, P-AIRCARS takes on the tasks of both doing the parallelization as well as scheduling. In an HPC environment the scheduling is usually done by a dedicated job scheduler, like Portable Batch System \citep[PBS,][]{pbs} or Slurm \citep{SLURM}.  Work is in progress to adapt P-AIRCARS for a cluster environment by incorporating an interface to a job scheduler. In parallel, we are also exploring the possibility of adapting P-AIRCARS for cloud computing platforms like Amazon Web Services (AWS), and Google Cloud Platform (GCP).

\section{Conclusion}\label{sec:conclusion}
P-AIRCARS represents the state-of-the-art pipeline for high-fidelity high dynamic-range spectro-polarimetric snapshot solar imaging at low radio frequencies. This work describes the implementation of the robust polarization calibration algorithm developed by \citet{Kansabanik2022_paircarsI}, along with several improvements to total intensity calibration, originally implemented in AIRCARS \citep{Mondal2019}. P-AIRCARS benefits from the experience gained and issues encountered during the extensive usage of AIRCARS, making it more robust. It is also much more user-friendly than AIRCARS. It delivers solar radio images with residual instrumental polarization leakages comparable to those achieved by high-quality MWA observations of non-solar fields \citep[e.g.][]{lenc2017,lenc2018}. Solar radio imaging has usually been the domain of specialists. Despite the usefulness of solar radio imaging being well established and the increasing availability of large volumes of excellent data in the public domain, the steep learning curve involved has been a hurdle in the large-scale use of these data. By providing a robust tool that dramatically reduces the effort involved in making high-quality solar radio images, we hope to help solar radio imaging become more mainstream.

The current implementation of P-AIRCARS is optimized for the MWA, however, the underlying algorithm is equally applicable to all centrally condensed arrays, including the upcoming SKA. The SKA is expected to be a discovery machine in the field of solar radio and heliospheric physics. P-AIRCARS and its predecessor, AIRCARS, are already leading to explorations of previously inaccessible phase spaces. They have enabled multiple interesting scientific results spanning a large range of solar phenomena using a SKA precursor, the MWA. We expect P-AIRCARS to form the workhorse for solar and heliospheric radio physics with the MWA and the stepping stone for the solar radio imaging pipeline for the SKA. 
  
\facilities{Murchison Widefield Array (MWA) \citep{lonsdale2009,Tingay2013,Wayth2018}}

\software{astropy \citep{price2018astropy}, matplotlib \citep{Hunter:2007}, Numpy \citep{Harris2020}, SciPy \citep{Scipy2020}, CASA \citep{mcmullin2007}, 
GSL \citep{Galassi2018}, WSClean \citep{Offringa2014}, Docker \citep{docker2014}, P-AIRCARS \citep{paircars_zenodo}.}

%\begin{acknowledgments}
\vspace{0.5cm}
\noindent This scientific work makes use of the Murchison Radio-astronomy Observatory (MRO), operated by the Commonwealth Scientific and Industrial Research Organisation (CSIRO). We acknowledge the Wajarri Yamatji people as the traditional owners of the Observatory site. Support for the operation of the MWA is provided by the Australian Government's National Collaborative Research Infrastructure Strategy (NCRIS), under a contract to Curtin University administered by Astronomy Australia Limited. We acknowledge the Pawsey Supercomputing Centre, which is supported by the Western Australian and Australian Governments. D. K. gratefully acknowledges Barnali Das (University of Delaware, Newark, USA) for useful discussions, and suggestions and also for providing a beautiful name for the pipeline. D. K. acknowledges Soham Dey (NCRA-TIFR, India) for testing this pipeline.  D. K., D. O., and A. B. acknowledge the support of the Department of Atomic Energy, Government of India, under project no. 12-R\&D-TFR-5.02-0700. S. M. acknowledges partial support by USA NSF grant AGS-1654382 to the New Jersey Institute of Technology. We also gratefully acknowledge the helpful comments from the anonymous referee, which have helped to improve the clarity and presentation of this work.
%\end{acknowledgments}

\bibliography{implementation}{}
\bibliographystyle{aasjournal}

\end{document}